\documentclass[runningheads]{llncs}
\usepackage[dvipdfmx]{graphicx}
\usepackage{multirow}
\usepackage{color}

\begin{document}

\title{Deep Factor Model\\ \small --Explaining Deep Learning Decisions for Forecasting Stock Returns with Layer-wise Relevance Propagation--}
\titlerunning{Deep Factor Model}
\author{Kei Nakagawa\inst{1,3}\orcidID{0000-0001-5046-8128} \and
Takumi Uchida\inst{3}\orcidID{0000-0003-4686-638X}
\and
Tomohisa Aoshima\inst{2,4}\orcidID{0000-0001-7810-4975}
}
\authorrunning{Nakagawa et al.}

\institute{Nomura Asset Management Ltd., Japan \email{kei.nak.0315@gmail.com}
\and Fujitsu Cloud Technologies Limited., Japan
\and University of Tsukuba, Graduate School of Business Sciences, Japan 
\and University of Tsukuba, Department of Risk Engineering, Japan}

\maketitle
\begin{abstract}

We propose to represent a return model and risk model in a unified manner with deep learning, which is a representative model that can express a nonlinear relationship. 
Although deep learning performs quite well, it has significant disadvantages such as a lack of transparency and limitations to  the interpretability of the prediction. 
This is prone to practical problems in terms of accountability.
Thus, we construct a multifactor model by using interpretable deep learning.
We implement deep learning as a return model to predict stock returns with various factors. 
Then, we present the application of layer-wise relevance propagation (LRP) to decompose attributes of the predicted return as a risk model. 
By applying LRP to an individual stock or a portfolio basis, we can determine which factor contributes to prediction. 
We call this model a deep factor model.
We then perform an empirical analysis on the Japanese stock market and show that our deep factor model has better predictive capability than the traditional linear model or other machine learning methods.
In addition , we illustrate which factor contributes to prediction.

\end{abstract}
\section{Introduction}

An essential tool of quantitative portfolio management is the multifactor model.
The model explains the stock returns through multiple factors. 
A general multifactor model in the academic finance field is sometimes used synonymously with the arbitrage pricing theory (APT) advocated by Ross~\cite{ross1976}.
The APT multifactor model includes a method of providing macroeconomic indicators a priori to explain stock returns and a method of extracting factors by factor analysis from past stock returns.

However, in practice, the Fama-French approach and the BARRA approach based on ICAPM~\cite{merton1973} are widely used.
The Fama-French or Barra multifactor models correspond to a method of finding stock returns using the attributes of individual companies such as investment valuation ratios represented by PER and PBR .

The Fama-French approach was introduced for the first time by Fama and French~\cite{fama1992}.
The Barra approach was introduced by Rosenberg~\cite{rosenberg1973} and was extended by Grinold and Kahn~\cite{grinold2000}.
It is calculated through cross-section regression analysis since it assumes that stock returns are explained by common factors.

In addition, there are two uses of the multifactor model. 
It can be employed both to enhance returns and to control risk.
In the first case, if one is able to predict the likely future value of a factor, a higher return can be achieved by constructing a portfolio that tilts toward ``good''   factors and away from ``bad'' ones.
In this situation, the multifactor model is called a return model or an alpha model.

On the other hand, by capturing the major sources of correlation among stock returns, one can construct a well-balanced portfolio that diversifies specific risk away. 
This is called a risk model.
There are cases where these models are confused when being discussed in the academic finance field.

For both the return model and the risk model, the relationship between the stock returns and the factors is linear in the traditional multifactor model mentioned above.
By contrast, linear multifactor models have proven to be very useful tools for portfolio analysis and investment management. The assumption of a linear relationship is quite restrictive. 
Considering the complexity of the financial markets, it is more appropriate to assume a nonlinear relationship between the stock returns and the factors.

Therefore, in this paper, we propose to represent a return model and risk model in a unified manner with deep learning, which is a representative model that can express a nonlinear relationship.
Deep learning is a state-of-the-art method for solving various challenging machine learning problems~\cite{goodfellow2016}, e.g., image classification, natural language processing, or human action recognition.
Although deep learning performs quite well, it has a significant disadvantage: a lack of transparency and limitations to the interpretability of the solution.
This is prone to practical problems in terms of accountability.
Thus, we construct a multifactor model by using interpretable deep learning.

We implement deep learning to predict stock returns with various factors as a return model.
Then, we present the application of layer-wise relevance propagation (LRP~\cite{bach2015}) to decompose attributes of the predicted return as a risk model.
LRP is an inverse method that calculates the contribution of inputs to the prediction made by deep learning.
LRP was originally a method for computing scores for image pixels and image regions to denote the impact of a particular image region on the prediction of a classifier for a particular test image.
By applying LRP to an individual stock or a quantile portfolio, we can determine which factor contributes to prediction.
We call the model a deep factor model.

We then perform an empirical analysis on the Japanese stock market and show that our deep factor model has better predictive power than the traditional linear model or other machine learning methods.
In addition, we illustrate which factor contributes to prediction.

\section{Related Works}
Stock return predictability is one of the most important issues for investors. 
Hundreds of papers and factors have attempted to explain the cross section of expected returns~\cite{subrahmanyam2010,mclean2016,harvey2016}.
Academic research has uncovered a large number of such factors, 314 according to Harvey et al.~\cite{harvey2016}, with the majority being identified during the last 15 years.

The most popular factors of today (Value, Size, and Momentum) have been studied for decades as part of the academic asset pricing literature and practitioner risk factor modeling research.
One of the best-known efforts in this field came from Fama and French in the early 1990s.
Fama and French~\cite{fama1992} put forward a model explaining US equity market returns with three factors: the market (based on the traditional CAPM  model), the size factor (large vs. small capitalization stocks), and the value factor (low vs. high book to market). 
The Fama-French three-factor model, which today includes Carhart's momentum factor~\cite{carhart1997}, has become a canon within the finance literature. 
More recently, the low risk~\cite{blitz2007} and quality factors~\cite{novy2013} have become increasingly well accepted in the academic literature.
In total, five factors are studied the most widely~\cite{jurczenko2015}.

Conversely, the investors themselves must decide how to process and predict returns, including the selection and weighting of such factors.
One way to make investment decisions is to rely upon the use of machine learning.
This is a supervised learning approach that uses multiple factors to explain stock returns as input values and future stock returns as output values. 
Many studies on stock return predictability using machine learning.
 have been reported .
Cavalcante et al.~\cite{cavalcante2016} presented a review of the application of several machine learning methods in financial applications.
In their survey, most of these were forecasts of stock market returns; however, forecasts of individual stock returns using the neural networks dealt with in this paper were also conducted.

In addition, Levin~\cite{levin1996} discussed the use of multilayer feed forward neural networks for predicting a stock return with the framework of the multifactor model.
To demonstrate the effectiveness of the approach, a hedged portfolio consisting of equally capitalized long and short positions was constructed, and its historical returns were benchmarked against T-bill returns and the S\&P500 index. 
Levin achieved persistent returns with very favorable risk characteristics. 

Abe and Nakayama~\cite{abe2018} extended this model to deep learning and investigated the performance of the method in the Japanese stock market.
They showed that deep neural networks generally outperform shallow neural networks, and the best networks also outperform
representative machine learning models. 
These results indicate that deep learning has promise as a skillful machine learning method to predict stock returns in the cross section.

However, these related works are only for use as a return model, and the problem is that the viewpoint as a risk model is lacking .

\section{Methodology -- Deep Factor Model}
\subsection{Deep Learning}
The fundamental machine learning problem is to find a predictor $f(x)$ of an output $Y$ given an input $X$.
As a form of machine learning, deep learning trains a model on data to make predictions, 
but it is distinguished by passing learned features of data through different layers of abstraction. 
Raw data is entered at the bottom level, and the desired output is produced at the top level,
which is the result of learning through many levels of transformed data. 
Deep learning is hierarchical in the sense that in every layer, the algorithm extracts features into factors, and a deeper level's factors become the next level's features.

A deep learning architecture can be described as follows.
We use $l \in  {1,...,L}$ to index the layers from $1$ to $L$, which are called hidden layers. 
The number of layers L represents the depth of our architecture.
We let $z^{(l)}$ denote the $l$-th layer, and so $X = z^{(0)}$.
The final output is the response $Y$ , which can be numeric or categorical. 

The explicit structure of a deep prediction rule is then
\begin{eqnarray}
	z^{(1)} &=& f^{(1)}(W^{(0)}X + b^{(0)}) \\
	z^{(2)} &=& f^{(2)}(W^{(1)}z^{(1)} + b^{(1)}) \\
	 && ... \nonumber \\ 
	Y & = & f^{(L)}(W^{(L-1)}z^{(L-1)} + b^{(L-1)}) 
\label{eq:deeplearning}
\end{eqnarray}
Here, $W^{(l)}$ are weight matrices, and $b^{(l)}$ are the threshold or activation levels.
$z^{(l)}$ are hidden features that the algorithm extracts. 
Designing a good predictor depends crucially on the choice of univariate activation functions $f^{(l)}$.
Commonly used activation functions are sigmoidal (e.g., $\frac{1}{(1 + exp(-x))}$, $cosh(x)$, or $tanh(x)$) or rectified linear units (ReLU) $max\{x, 0\}$. 

\subsection{Layer-Wise Relevance Propagation}
\begin{figure}
\includegraphics[width=0.9\textwidth]{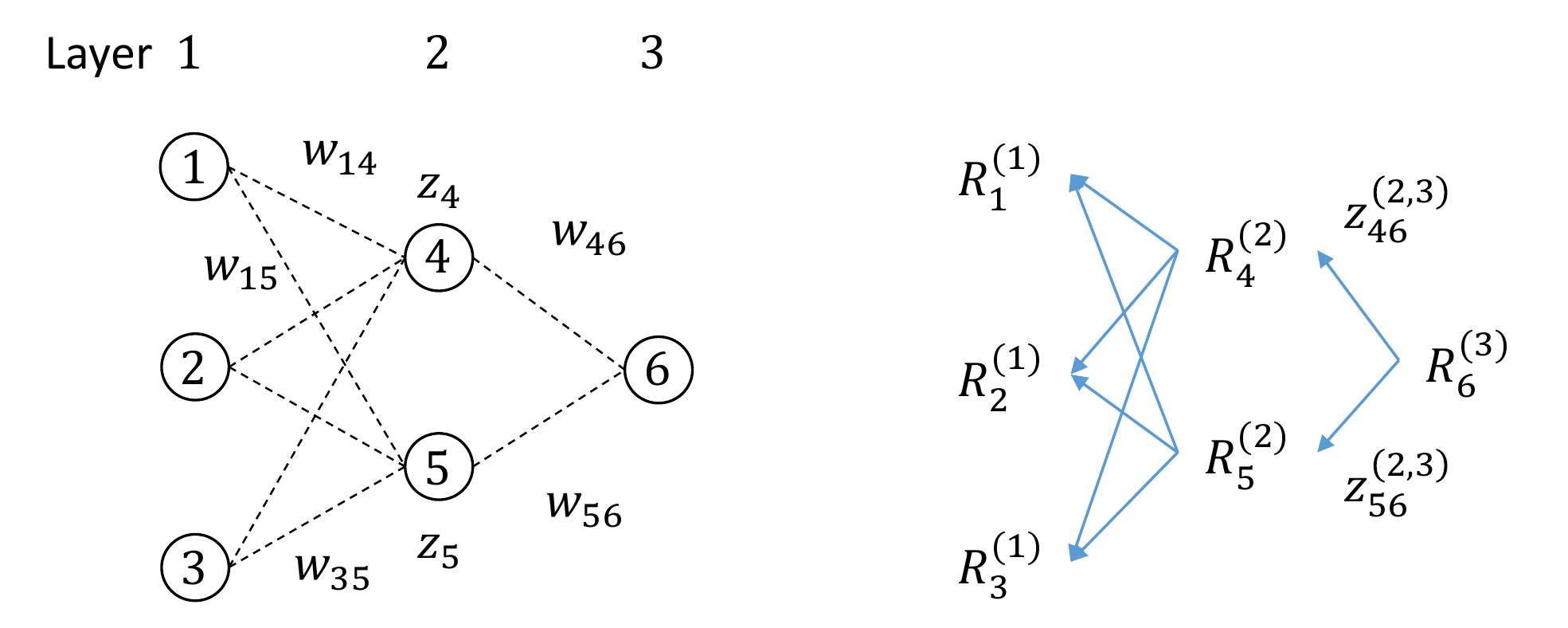}
\caption{LRP with toy example} \label{fig:LRP}
\end{figure}

LRP is an inverse method that calculates the contribution of the prediction made by the network.
The overall idea of decomposition is explained in~\cite{bach2015}. 
Here, we briefly reiterate some basic concepts of LRP with a toy example (Fig. \ref{fig:LRP}).
Given input data $x$, a predicted value $f(x)$ is returned by the model denoted as function $f$.
Suppose the network has $L$ layers, each of which is treated as a vector with dimensionality $V(l)$, 
where $l$ represents the index of layers. 
Then, according to the conservation principle, LRP aims to find a relevance score $R_d$ for each vector element in layer $l$ such that the following equation holds:

\begin{equation}
	f(x) = \sum_{d \in V(L)} R_{d}^{(L)} = ... = \sum_{d \in V(l)} R_d^{(l)} = ... = \sum_{d \in V(1)} R_d^{(1)}
    \label{eq:conservation}
\end{equation}

As we can see in the above formula (\ref{eq:conservation}), LRP uses the prediction score as the sum of relevance scores for the last layer of the network, and maintains this sum throughout all layers.

Fig. \ref{fig:LRP} shows a simple network with six neurons. 
$w_{ij}$ are weights, $z_i$ are outputs from activation, and $R_i^{(l)}$ are relevance scores to be calculated.
Then, we have the following equation:

\begin{equation}
	f(x) = R_{6}^{(3)} =  R_{5}^{(2)} + R_{4}^{(2)} = R_{3}^{(1)} + R_{2}^{(1)} + R_{1}^{(1)}
    \label{eq:toy}
\end{equation}

Furthermore, the conservation principle also guarantees that the inflow of relevance scores to one neuron equals the outflow of relevance scores from the same neuron.
$z_{ij}^{(l,l+1)}$ is the message sent from neuron $j$ at layer $l + 1$ to neuron $i$ at layer $l$.
In addition, $R_d^{(l)}$ is computed using network weights according to the equation below:

\begin{equation}
	R_i^{(l)} = \sum_j \frac{z_{ij}^{(l,l+1)}}{\sum_{k} z_{kj}^{(l,l+1)}}R_j^{(l+1)},z_{ij}^{(l,l+1)} = w_{ij} z_{i}^{(l)} 
\end{equation}

Therefore, LRP is a technique for determining which features in a particular input vector contribute most strongly to a neural network's output.

\subsection{Deep Factor Model}
In this paper, we propose to represent a return model and risk model in a unified manner with deep learning, which is a representative model that can express a nonlinear relationship.
We call the model a deep factor model.
First, we formulate a nonlinear multifactor model with deep learning as a return model.

The traditional fundamental multifactor model assumes that the stock return $r_i$ can be described by a linear model:

\begin{equation}
	r_i = \alpha_i + X_{i1}F_1+...+ X_{iN}F_N + \varepsilon_i
    \label{eq:factormodel}
\end{equation}
where $F_i$ are a set of factor values for stock $i$, $X_{in}$ denotes the exposure to factor $n$, $\alpha_i$ is an intercept term that is assumed to be equal to a risk-free rate of return under the APT  framework, and $\varepsilon_i$ is a random term with mean zero and is assumed to be uncorrelated across other stock returns.
Usually, the factor exposure $X_{in}$ is defined by the linearity of several descriptors .

While linear multifactor factor models have proven to be very effective tools for portfolio analysis and investment management, the assumption of a linear relationship is quite restrictive. 
Specifically, the use of linear models assumes that each factor affects the return independently. Hence, 
they ignore the possible interaction between different factors. 
Furthermore, with a linear model, the expected return of a security can grow without bound as its exposure to a factor increases. 

Considering the complexity of the financial markets, it is more appropriate to assume a nonlinear relationship between the stock returns and the factors.
Generalizing (\ref{eq:factormodel}), maintaining the basic premise that the state of the world can be described by a vector of factor values and that the expected stock return is determined through its coordinates in this factor world leads to the nonlinear model:

\begin{equation}
	r_i = \tilde{f}(X_{i1} ,..., X_{iN}, F_1 ,..., F_N) + \varepsilon_i
\label{eq:nlfactormodel}
\end{equation}
where $\tilde{f}$ is a nonlinear function.

The prediction task for the nonlinear model (\ref{eq:nlfactormodel}) is substantially more complex than that in the linear case since it requires both the estimation of future factor values as well as a determination of the unknown function $\tilde{f}$. 
As in a previous study~\cite{levin1996}, the task can be somewhat simplified if factor estimates are replaced with their historical means $\bar{F}_n$.
Since the factor values are no longer variables, they are constants.
For the nonlinear model (\ref{eq:nlfactormodel}), the expression can be transformed as follows:

\begin{eqnarray}
	r_i &=& \tilde{f}(X_{i1} ,..., X_{iN}, \bar{F}_1,...,\bar{F}_N) + \varepsilon_i \\
		&=& f(X_{i1} ,..., X_{iN}) + \varepsilon_i         
\label{eq:deepfactormodel}
\end{eqnarray}
where $X_{in}$ is now the security's factor exposure at the beginning of the period over which we wish to predict.
To estimate the unknown function $f$, a family of models needs to be selected, from which a model is to be identified. 
In the following, we propose modeling the relationship between factor exposures and future stock returns using a class of deep learning.

\begin{figure}
\begin{center}
\includegraphics[width=0.9\textwidth]{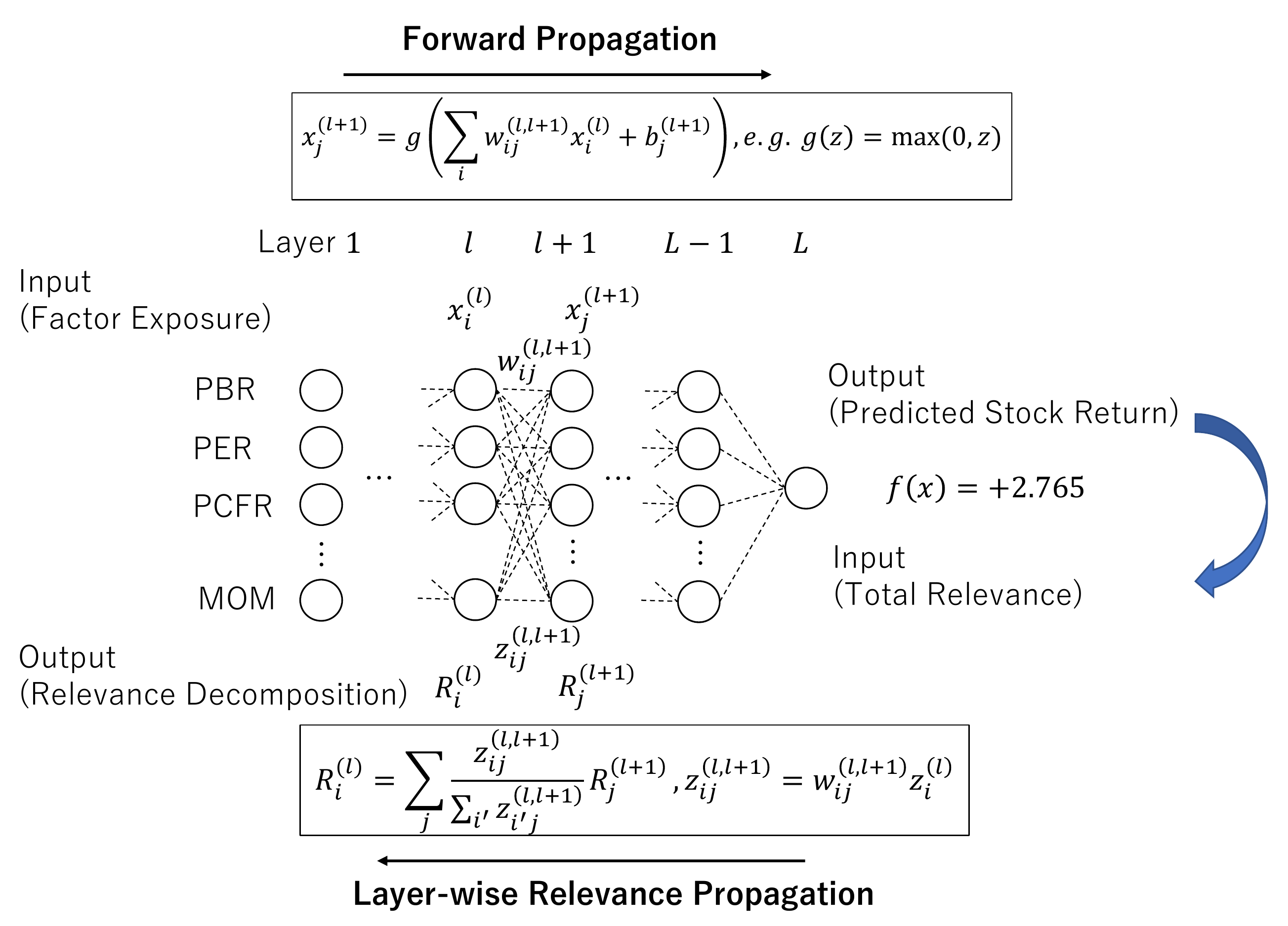}
\caption{Deep factor model} \label{fig:deepfactor}
\end{center}
\end{figure}

However, deep learning has significant disadvantages such as a lack of transparency and limitations to the interpretability of the solution. 
This is prone to practical problems in terms of accountability.
Then, we present the application of LRP to decompose attributes of the predicted return as a risk model.
By applying LRP to an individual stock or a quantile portfolio, we can determine which factor contributes to prediction.
If you want to show the basis of the prediction for a stock return, you can calculate LRP using the inputs and outputs of the stock.
In addition, in order to obtain the basis of prediction for a portfolio, calculate LRPs of the stocks included in that portfolio and take their average.
Then, by aggregating the factors, you can see which factor contributed to the prediction.
Fig. \ref{fig:deepfactor} shows an overall diagram of the deep factor model.

\section{Experiment on Japanese Stock Markets}
\subsection{Data}
We prepare a dataset for TOPIX index constituents.
TOPIX is a well-accepted stock market index for the Tokyo Stock Exchange (TSE) in Japan, tracking all domestic companies of the exchange's First Section. It is calculated and published by the TSE. 
As of March 2016, the index is composed of 1,948 constituents .
The index is also often used as a benchmark for overseas institutional investors who are investing in Japanese stocks. 

We use the 5  factors and 17 factor exposures listed in Table \ref{factor}.
These are used relatively often in practice and are studied the most widely in academia~\cite{jurczenko2015}.

In calculating these factors, we acquire necessary data from the Nikkei Portfolio Master and Bloomberg.
Factor exposures are calculated on a monthly basis (at the end of month) from December 1990 to March 2016 as input data.
Stock returns with dividends are acquired on a monthly basis (at the end of month) as output data.

\begin{table}[htbp]
\centering
\caption{Factors and factor descriptors}
\label{factor}
\scalebox{0.9}{\begin{tabular}{|l|l|l|}
\hline
\multicolumn{1}{|c|}{Factor} & \multicolumn{1}{c|}{Descriptors} & \multicolumn{1}{c|}{Formula}                                                          \\ \hline
\multirow{3}{*}{Risk}        & 60VOL                              & Standard deviation of stock returns in the past 60 months                             \\ \cline{2-3} 
                             & BETA                               & Regression coefficient of stock returns and market risk premium \\ \cline{2-3} 
                             & SKEW                               & Skewness of stock returns in the past 60 months                                       \\ \hline
\multirow{4}{*}{Quality}     & ROE                                & Net income/Net Assets                                                                 \\ \cline{2-3} 
                             & ROA                                & Operationg Profit/Total Assets                                                        \\ \cline{2-3} 
                             & ACCRUALS                           & Operating Cashflow ‐ Operationg Profit                                                \\ \cline{2-3} 
                             & LEVERAGE                           & Total Liabilities / Total Assets                                                      \\ \hline
\multirow{3}{*}{Momentum}    & 12-1MOM                            & Stock returns in the past 12 months except for past month                               \\ \cline{2-3} 
                             & 1MOM                               & Stock returns in the past month                                                     \\ \cline{2-3} 
                             & 60MOM                              & Stock returns in the past 60 months                                                   \\ \hline
\multirow{5}{*}{Value}       & PSR                                & Sales/Market Value                                                                    \\ \cline{2-3} 
                             & PER                                & Net Income / Market Value                                                             \\ \cline{2-3} 
                             & PBR                                & Net Assets / Market Value                                                             \\ \cline{2-3} 
                             & PCFR                               & Operating Cashflow / Market Value                                                     \\ \hline
\multirow{2}{*}{Size}        & CAP                                & log(Market Value)                                                                     \\ \cline{2-3} 
                             & ILLIQ                              & average(Stock Returns/Trading Volume)                                                 \\ \hline
\end{tabular}}
\end{table}

\subsection{Model}
Our problem is to find a predictor $f(x)$ of an output $Y$, next month's stock returns given an input $X$, and various factors. 
One set of training data is shown in Table \ref{data}.
In addition to the proposed deep factor model, we use a linear regression model as a baseline, and support vector regression(SVR) and random forest as comparison methods.
The deep factor model is implemented with TensorFlow~\cite{abadi2016}, and the comparison methods are implemented with scikit-learn~\cite{pedregosa2011}.
Table \ref{method} lists the details of each model.

\begin{table}[htbp]
\centering
\caption{Details of each method}
\label{method}
\scalebox{0.9}{\begin{tabular}{|l|l|l|}
\hline
\multicolumn{2}{|l|}{Model}                  & Description                                                                                                                                                                                                  \\ \hline
\multirow{2}{*}{Deep Factor Model} & Model 1 & \begin{tabular}[c]{@{}l@{}}The hidden layers are \{80-50-10\}.\\ We use the ReLU as the activation function \\ and Adam~\cite{kinga2015} for the optimization algorithm.\end{tabular} \\ \cline{2-3} 
                                   & Model 2 & \begin{tabular}[c]{@{}l@{}}The hidden layers are \{80-80-50-50-10-10\}.\\ We use the ReLU as the activation function \\ and Adam~\cite{kinga2015} for the optimization algorithm.\end{tabular}          \\ \hline
\multicolumn{2}{|l|}{Linear Model}           & \begin{tabular}[c]{@{}l@{}}Linear models is implemented with scikit-learn \\ with the class ``sklearn.linear\_model.LinearRegression''\\ All parameters are default values in this class.\end{tabular}            \\ \hline
\multicolumn{2}{|l|}{SVR}                    & \begin{tabular}[c]{@{}l@{}}Support vector regression (SVR) is implemented \\ with scikit-learn with the class ``sklearn.svm.SVR''.\\ All parameters are default values in this class.\end{tabular}                \\ \hline
\multicolumn{2}{|l|}{Random Forest}          & \begin{tabular}[c]{@{}l@{}}Random Forest is implemented with scikit-learn \\ with the class ``sklearn.ensemble.RandomForestRegressor''.\\ All parameters are default values in this class.\end{tabular}           \\ \hline
\end{tabular}}
\end{table}

We train all models by using the latest 60 sets of training data from the past 5 years. 
The models are updated by sliding one month ahead and carrying out a monthly forecast.
The prediction period is 10 years, from April 2006 to March 2016 (120 months). 
In order to verify the effectiveness of each method, we compare the prediction accuracy of these models and the profitability of the quintile portfolio.
We construct a long/short portfolio strategy for a net-zero investment to buy top stocks and to sell bottom stocks with equal weighting in quintile portfolios.
For the quintile portfolio performance, we calculate the annualized average return, risk, and Sharpe ratio.
In addition, we calculate the average mean absolute error (MAE) and root mean squared error (RMSE) for the prediction period as the prediction accuracy .

\begin{table}[htbp]
\centering
\caption{One set of training data for March 2016.}
\label{data}
\scalebox{0.9}{\begin{tabular}{|l|l|}
\hline
Input: 80 dim                                                                                                                                                               & Output: 1 dim                                                                                   \\ \hline
\multicolumn{1}{|c|}{\begin{tabular}[c]{@{}c@{}}Factor Descriptors: \\ 16 × 5 dim\\ \\ February  2016\\ November 2015\\ August 2015\\ May 2015\\ February 2015\end{tabular}} & \multicolumn{1}{c|}{\begin{tabular}[c]{@{}c@{}}Return:\\ 1 dim\\ \\ \\ March 2016\end{tabular}} \\ \hline
\end{tabular}}
\end{table}

\subsection{Results}

Table \ref{results} lists the average MAE and RMSE of all years and the annualized return, volatility, and Sharpe ratio for each method.
In the rows of the table, the best number appears in bold.
Deep factor model 1 (shallow) has the best prediction accuracy in terms of MAE and RMSE as in the previous study~\cite{levin1996,abe2018}.
On the other hand, deep factor model 2 (deep) is the most profitable in terms of the Sharpe Ratio.
The shallow model is superior in accuracy, while the deep one is more profitable.
In any case, we find that both models 1 and 2 exceed the baseline linear model, SVR, and random forest in terms of accuracy and profitability.
These facts imply that the relationship between the stock returns in the financial market and the factor is nonlinear, rather than linear.
In addition, a model that can capture such a nonlinear relationship is thought to be superior.

\begin{table}[htbp]
\centering
\caption{Average MAE and RMSE of all years and annualized return, volatility, and Sharpe ratio for each method.}
\label{results}

\scalebox{0.9}{\begin{tabular}{|l|c|c|c|c|c|}
\hline
\multirow{2}{*}{}   & \multicolumn{2}{c|}{Deep Factor Model} & \multirow{2}{*}{Linear Model} & \multirow{2}{*}{SVR} & \multirow{2}{*}{Random Forest} \\ \cline{2-3}
                    & Model 1            & Model 2           &                               &                      &                                \\ \hline
Return {[}\%{]}     & \bf{10.81}              & 10.31             & 8.17                          & 1.46                 & 0.12                           \\ \hline
Volatility {[}\%{]} & 7.65               & 6.86              & 8.20                          & 9.66                 & \bf{5.43}                           \\ \hline
Sharpe Ratio        & 1.41               & \bf{1.50}              & 1.00                          & 0.15                 & 0.02                           \\ \hline
MAE                 & \bf{0.0663}             & 0.0669            & 0.0679                        & 0.1713               & 0.0728                         \\ \hline
RMSE                & \bf{0.0951}             & 0.0953            & 0.0965                        & 0.1962               & 0.1024                         \\ \hline
\end{tabular}}
\end{table}

\subsection{Interpretation}

Here, we try to interpret the stock of the highest predicted stock return and the top quintile portfolio based on the factor using deep factor model 2 as of the last time point of February 2016. 

\begin{figure}
\centering
\includegraphics[width=0.9\textwidth]{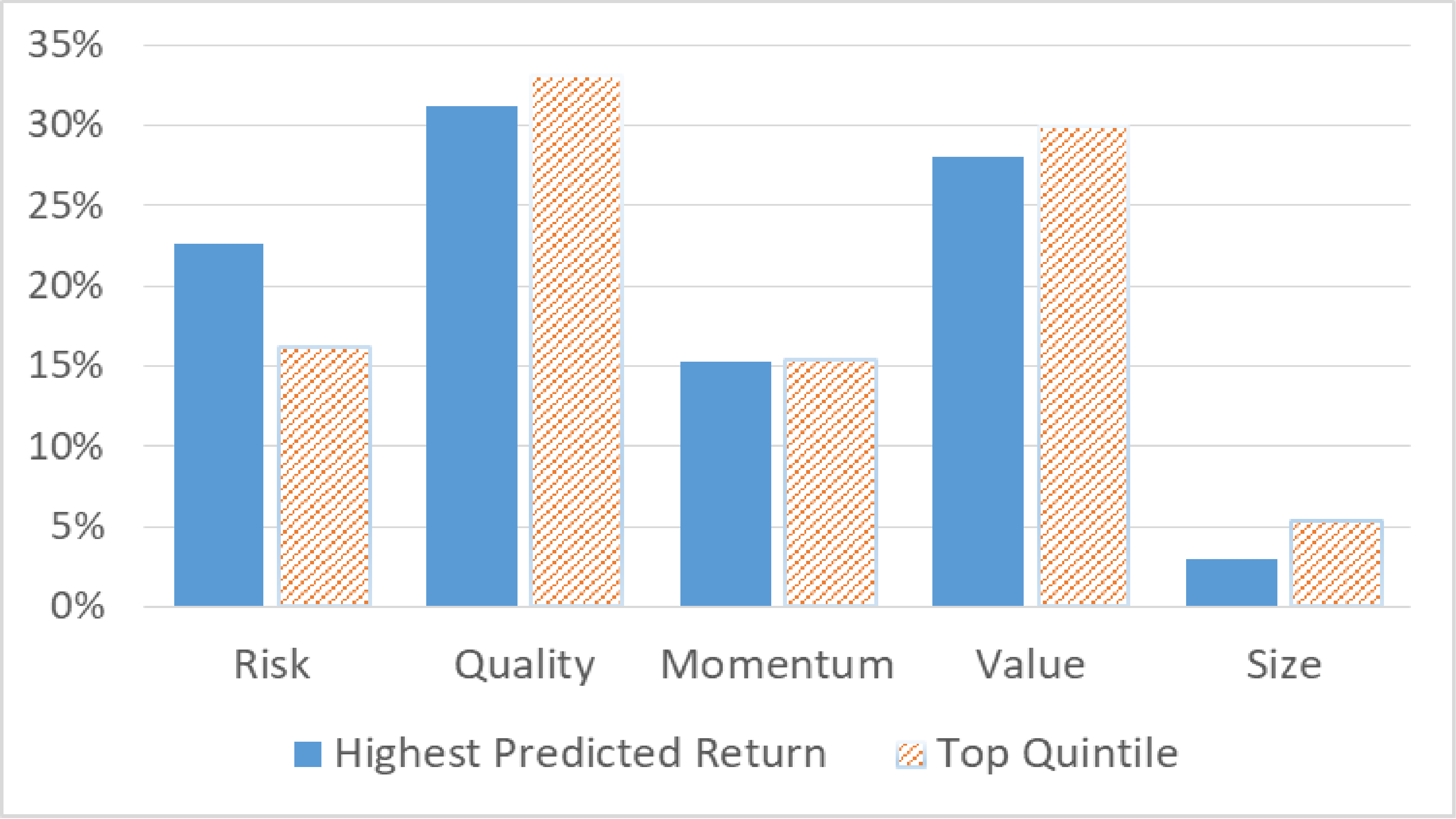}
\caption{Interpreting highest predicted return and top quintile portfolio based on factor using network as of last time point of February 2016} \label{fig:LRP_factor}
\end{figure}

Fig. \ref{fig:LRP_factor} shows which factor contributed to the prediction in percentages using LRP.
The contributions of each descriptor calculated by LRP are summed for each factor and are displayed as a percentile.
We observe that the quality and value factors account for more than half of the contribution to both the stock return and quintile portfolio.
In general, the momentum factor is not very effective, but the value factor is effective in the Japanese stock markets~\cite{fama2012}.
On the other hand, there is a significant trend in Japan to evaluate companies that will increase ROE  over the long term because of the appearance of the Corporate Governance Code\footnote{The Corporate Governance Code is designed to encourage companies to set ROE targets, increase the number of external directors, and unwind cross-shareholdings. 
This went live on 1st June 2015. Before year’s end, each listed company will publish a corporate governance document outlining its policy, which will become an annual publication going forward.}.
In response to this trend, the quality factor including ROE is gaining attention.
Moreover, the contribution of the size factor is small, and it turns out that there is a widely profitable opportunity regardless of whether the stock is large or small.

Next, we quantitatively verify the risk model by LRP.
Table \ref{tbl:LRP_COR} shows the correlation coefficients between each factor and the predicted return in the top quintile portfolio.
The correlation coefficients are calculated by averaging the correlation coefficients between each descriptor and the predicted return by each factor. The influence of the value and size factor differs when looking at LRP and correlation.
The value factor has a large contribution to LRP and a small contribution to the correlation coefficients. The size factor has the opposite contributions.
Therefore, without LRP, we could misinterpret the return factors.

\begin{table}[htbp]
\centering
\caption{Correlation coefficients between each factor and predicted return in top quintile portfolio.}
\scalebox{0.9}{\begin{tabular}{|c|l|l|l|l|l|l|}
\hline
\multicolumn{2}{|l|}{}          & Risk & Quality & Momentum & Value & Size \\ \hline
\multirow{2}{*}{correlation} & spearman & 0.14     &  0.22       &   0.24       &   0.08    & 0.14     \\ \cline{2-7} 
                     & kendall  & 0.10     &   0.15      &   0.17       &   0.06    & 0.10     \\ \hline
\end{tabular}}
\label{tbl:LRP_COR}
\end{table}

\section{Conclusion}
We presented a method by which deep-learning-based models can be used for stock selection and risk decomposition. 
Our conclusions are as follows:
\begin{itemize}
 \item  The deep factor model outperforms the linear model. This implies that the relationship between the stock returns in the financial market and the factors is nonlinear, rather than linear. The deep factor model also outperforms other machine learning methods including SVR and random forest.
 \item  The shallow model is superior in accuracy, while the deep model is more profitable.
 \item  Using LRP, it is possible to intuitively determine which factor contributed to prediction.
\end{itemize}

For further study, we would like to expand our deep factor model to a model that exhibits dynamic temporal behavior for a time sequence such as RNN.

\end{document}